\documentstyle[12pt]{article}

\def\be{\begin {equation}}

\def\ee{\end {equation}}

\def\bea{\begin{eqnarray}}

\def\eea{\end{eqnarray}}

\begin{document}

\vspace{5cm}

\thispagestyle{empty}

\begin{center}

{\LARGE \bf Unified Approach to KdV Modulations}

\vspace{0.5cm}
{\Large Gennady A. El}

Institute of Terrestrial Magnetism, Ionosphere and Radio Wave
Propagation, Russian Academy of Sciences, Troitsk, Moscow Region

\vspace{0.5cm}
{\Large Alexander L. Krylov}

O.Yu. Shmidt Institute of Earth Physics, Russian Academy of
Sciences, Moscow

\vspace{0.5cm}
{\Large Stephanos Venakides}

Duke University, Durham, NC

\end {center}
\vspace{1cm}

{\bf Abstract}
We develop a unified approach to integrating the
Whitham modulation equations. Our approach is based on the formulation
of the initial value problem for the zero dispersion KdV
as the steepest descent for the scalar Riemann-Hilbert problem
\cite{DVZ} and on the method of generating
differentials for the KdV-Whitham hierarchy \cite{El}.
By assuming the hyperbolicity of the zero-dispersion limit
for the KdV with general initial data, 
we bypass the inverse scattering transform and
produce the symmetric system of algebraic
equations describing motion of the modulation parameters
plus the system of inequalities determining the number
the oscillating phases at any fixed point on the $x, t$ - plane.
The resulting system effectively solves the zero dispersion KdV
with an arbitrary initial data.

\section{Introduction}
The initial value problem for the Korteweg -- de Vries equation

\be \label{KdV}
u_t -6uu_x +\epsilon^2 u_{xxx} =0, \ \ \
u(x,0)=u_0 (x)
\ee
in the zero dispersion limit $\epsilon \to 0$ has been the
subject of intense studies during more than 25 years.
The physical interest of this limit is that it allows to model
the phenomenon of dispersive shocking in dissipationless
dispersive media. In contrast to the usual dissipative hydrodynamics,
the regularization of a shock occurs here through the generation
of small-scale nonlinear oscillations. Gurevich and Pitaevskii
in \cite{GP} proposed to describe these oscillations
with the aid of the one-phase Whitham modulation equations \cite{Wh}.
Multiphase analog of the Whitham equations was derived
by Flaschka, Forest and McLaughlin \cite{FFM}.
Later, Lax and Levermore \cite{LL}
rigorously showed
that the multiphased averaged  equations appear in the zero
dispersion limit of
the initial value problem for the KdV equation with asymptotically
reflectionless initial data. For the reflecting potentials the zero
dispersion
theory was constructed by Venakides \cite{V85} who also identified
the parameters
describing the weak limit of the solution in the Lax-Levermore
approach with the Riemann invariants of the modulation equations
in \cite{FFM} (see \cite{V85a}, \cite{V87}).

The modulation system itself, however, has to be integrated to
reveal dependence of the Riemann invariants (modulation
parameters) on $x$ and $t$. For this, the observation of
Tsarev \cite{Ts} (see also \cite{DN}) who generalized the
classical hodograph
method to a multidimensional (in the space of dependent
variables) case was crucial. The result of applying this
generalized hodograph transform to the
averaged KdV is the overdetermined consistent system of linear PDE's.
Tsarev's results were put into the algebro-geometrical
setting by Krichever \cite{Kr88}, \cite{Kr94}.

Each solution
to the Tsarev system gives rise to some local solution of  the
nonlinear modulation system. These solutions generically
exist only within definite regions  of the $x, t$ - plane.
To obtain the global
solution one should supplement the constructed local solutions
with the information about the number $N$ of nonlinear phases
in each region and provide the smooth matching of solutions
with different $N$ on the phase transition boundaries.
The simplest, yet very important, class of problems with
$N\le1$ has been investigated in works of Tian \cite{T1}, \cite{T2},
Gurevich,
Krylov, El  and their collaborators \cite{GKE}, \cite{GKME}
who obtained a number of exact solutions for the initial value
problems with monotone and hump-like data having the only break
point. Tian \cite{T2} was also able to prove that the solutions to the Whitham
system in this case do globally belong to $N=1$.

More general local solutions to the modulation system
in the case of arbitrary $N$ were constructed
in a symmetric form by  El \cite{El} with the aid of
fundamental solution (an analog of the Green function)
to the Tsarev equations.

Some global solutions involving the case $N=2$ were constructed
recently by Grava \cite{Gr} on the basis of Dubrovin's variational
approach to the Whitham equations \cite{D}.

The common feature of all pointed methods for constructing
the global solutions to the Whitham equations is the necessity
to follow all earlier times $t<t_0$ to obtain the
solution at $t_0$. This difficulty has been bypassed resently
by Deift, Venakides and Zhou \cite{DVZ} who obtained
the algebraic equations for the KdV modulations avoiding the
integration of the modulation equations themselves. The method
of DVZ is based on reformulation of the initial value problem for
the zero dispersion KdV with decaying solitonless
analytic initial data as the steepest descent for the
scalar Riemann-Hilbert problem for the complex phase function.
As a result, they obtain not only the local solutions to the
modulation equations, but also the system of inequalities
enabling to determine the number of phases
locally at each point of the $x, t$-plane.

It is clear, however, that the requirements of decaying and
pure reflection for the  initial potential are not essential
in the zero dispersion limit. Really,
assuming the finite speed of propagation
(i.e. validity of the Whitham equations) for the zero-dispersion KdV
with an
arbitrary initial data one can see that only the finite part of the initial
data contributes to the solution at any finite $t$. In this case,
considering the evolution of 'multihump' potential and and then
tending number of 'humps'
to infinity one arrives at the solution for nondecaying initial data.
Certainly, for the finite $\epsilon$ the results obtained in this way are
valid  within finite time interval until the semiclassical
(Whitham) spectrum begins to compete with the fine spectrum of the potential.
The assumption of hyperbolicity in
the limit studied allows also to omit the requirement of analyticity for the
initial data.

In this work, we combine methods of Deift, Venakides, and Zhou
\cite{DVZ} and El \cite{El} to
produce the global solution to the KdV modulation equations
with an arbitrary initial data. Namely, by {\it assuming}
validity of the modulation equations in the case
of the zero dispersion KdV with an arbitrary initial data
we reformulate the initial value problem for the Whitham system  as the
Riemann-Hilbert problem and produce the symmetric system of
algebraic equations supplemented by the system of inequalities determining
both the motion of the
Riemann invariants and the change of the genus of the Riemann surface
(number of oscillating phases).  We represent the local part of the
obtained general
solution in a potential form with the generalized functional of the
Peierls-Fr\"ohlich type \cite{Bel}, \cite{Kr}, \cite{BDK} as a potential.
Recently the functionals of this type in more particular form
were shown to give rise by minimization to the global solutions for the
Whitham equations with monotone initial data \cite{D}.

We also make
identification of the part of the results obtained in \cite{DVZ} with the
results obtained earlier in \cite{Kr88} , \cite{El} and \cite{D}
which unifies  different approaches developed independently in this area.

\section{Summary of the Riemann-Hilbert Steepest
Descent Method for the
Zero Dispersion KdV with Decaying Solitonless Initial Data}

It is well known that the inverse scattering transform
can be reformulated as a matrix Riemann-Hilbert (RH) problem
(see \cite{Sh}, \cite{BDT}).
Deift, Venakides, and Zhou \cite{DVZ} showed by considering
the quasiclassical asymptotics for the initial value
problem (\ref{KdV}) in the case of one-hump solitonless initial data
$0\le u_0(x)\le 1$ that this problem   can be
asymptotically (as $\epsilon \to 0$) reduced to the scalar
RH problem for the
complex phase function $g(\lambda)$.  We present here the resulting formulas
(for details see \cite{DVZ}, \cite{V}).

Let the interval $(0,1)$ of the real axis on the $\lambda$ -
plane is partitioned into a finite set of intervals
$\{G_k (k=0, 1, \dots, N): \ (0, r_1), (r_2, r_3), \dots, (r_{2N},
r_{2N+1}) \}$, $\{B_n (n=1, \dots, N+1) : \ (r_1,r_2),
(r_3, r_4), \dots, (r_{2N+1}, 1)\} $.

We introduce the combination
\be \label{alpha}
\alpha(\lambda) = 4t\lambda^{3/2}
+ x\lambda^{1/2} \, ,
\ee
which will be important hereafter, and the phases of the reflection
(from the right) and the transmission coefficients for the scattering
on the potential $u_0(x)$.

\be \label{rho+}
\rho_+(\lambda)= x^{+}\lambda ^{1/2} + \int \limits^{\infty}_{x^{+}}
[\lambda^{1/2}-(\lambda-u_0(x'))^{1/2}]dx' \, ,
\ee

\be \label{tau}
\tau(\lambda)= \int \limits^{x^{+}}_{x^{-}}(u_0(x')-\lambda)^{1/2} dx'\, .
\ee

\noindent where
$x^{\pm}=x^{\pm}(\lambda)$
are the roots of the equation $u_0(x)=\lambda$.

\noindent Then the RH problem defining the complex phase function
$g(\lambda)$, which plays the central role in the future, takes the form

\noindent On the intervals $\{G_k, (k=0,\dots,N): \ (0,r_1), (r_2,r_3),
\dots, (r_{2N}, r_{2N+1})\}$
the following relations hold:

\be \label{egaps0}
\frac{g_+' + g_-'}{2}= \rho'_+ -\alpha'
\ee

\be \label{igaps0}
-\tau<\frac{g_+ - g_-}{2i}<0 \, ,
\ee

It follows from the equality (\ref{egaps0}) that on each
interval $G_k$:  $g_+ +g_- - 2\rho_+
+2\alpha = -\Omega_k$, where $\Omega_k$ is some constant of integration;
without loss of generality one can put $\Omega_0=0$.

On the intervals $\{B_k, (k=1,\dots, N): \ (r_1,r_2),
(r_3,r_4), \dots, (r_{2N-1},r_{2N})\}$
we have
\be \label{ebands0}
\frac{g_+ -g_-}{2i}=-\tau \, ,
\ee

\be \label{ibands0}
\frac{g_+' + g_-'}{2}< \rho'_+ -\alpha'  \, .
\ee
The latter ineguality is due to the further reduction of the RH
problem with the aid of the steepest descent method \cite{DVZ}.

On the remaining interval $\{B_{N+1}:(r_{2N+1},1)\}$
there exist two possibilities:

\be \label{A0}
\hbox{case A:} \qquad
\frac{g_+ - g_-}{2i} =-{\tau} \, , \qquad
\frac{g_+' + g_-'}{2}< \rho'_+(\lambda) -\alpha'  \, .
\ee
which coincides with the conditions (\ref{ebands0}), (\ref{ibands0})

\noindent or

\be \label{B0}
\hbox{case B:} \ \qquad
g_+ - g_- =0 \, , \qquad
\frac{g_+' + g_-'}{2}> \rho'_+(\lambda) -\alpha'  \, .
\ee
Also we have outside the interval $(0,1)$
\be \label{<00}
g_+ +g_- =0 \qquad  \hbox{if} \qquad \lambda<0,
\ee
\be \label{>00}
g_+ - g_- =0 \qquad \hbox{if} \qquad \lambda>1 \, .
\ee

In all above formulas we denoted

$$
g_{\pm}(\lambda)\equiv \lim \limits_{\delta\to 0}g(\lambda \pm i\delta) \,
$$

The additional requirement imposed on the function $g(\lambda)$
is
\be \label{}
\hbox{The functions} \ (\sqrt{\lambda}g'(\lambda))_{\pm}
\ \ \hbox{are continuous for real }  \lambda \, ,
\ee
Also it follows from \cite{DVZ} that the following asymptotic holds
\be \label{gasymp}
g(\lambda)=g_1/ \lambda^{1/2} + O(1/\lambda) \qquad
\hbox{as} \ \lambda \to \infty \, .
\ee
One can see that just formulated RH problem contains not only
equations defining the function $g(\lambda)$ on the intervals
$B_k$ and $G_k$ but also the inequalities which define
the overall number $2N$ of the intervals (if $N$ has been chosen
wrongly then at least one of the  inequalities will be violated).

We observe that for each fixed $x$ and $t$, $g'$ also satisfies a scalar RH
problem on the real axis. Indeed, $g_+'+g_-'=0$ when $\lambda <0$, and
$g_+'-g_-'=0$ when $\lambda >1$.  When $0<\lambda <1$, the equalities
(\ref{egaps0}), (\ref{ebands0}), (\ref{A0}) and (\ref{B0}) specify
$g_+'+g_-'-2\rho'+2\alpha'=0$ on each interval $G_k$, while $g_+'-g_-'$ either
equals 0 or equals $-2i\tau'$ outside these intervals.
We only consider $g_+'-g_-'=-2i\tau'$ when $\lambda$ lies
between any two of the $G_k$'s while on the remaining interval
$(r_{2N+1}, 1)$, we examine both possibilities i.e.
$g_+'-g_-'= -2i\tau'$ (case A) or $g_+'-g_-'= 0$ (case B).

Necessarily $g'(\lambda )$ has the following form,

\be \label{g'}
g'(\lambda )=\sqrt{R_{2N+1}(\lambda)}\left(\int_{\cup G_k}\frac{2\rho'(\mu)
-2\alpha'(\mu)}{
\sqrt{R_{2N+1}^+(\mu)}(\mu-\lambda)}\frac{d\mu}{2\pi i}
+\int_{(0,E)\setminus\cup G_k}
\frac{-2i\tau'(\mu)}{\sqrt{R_{2N+1}^+(\mu)}(\mu-\lambda)}\frac{d\mu}{2\pi
i},\right) \, , \ee
where
\be \label{R}
R_{2N+1}(\lambda)=\prod_{j=1}^{2N+1}
(\lambda-r_j) \, ,
\ee
and $E=1$ in the case A and $E=r_{2N+1}$ in the case B. Here,  $\sqrt{R_{2N+1}
(\lambda )}$ is positive for $\lambda >r_{2N+1}$.  Also,
$\sqrt{R_{2N+1}^+(\lambda )}$ denotes the boundary value from above.

A sufficient number of conditions to determine the endpoints of the intervals
$G_j$ can now be written down. Indeed, the condition
$g(\lambda )=O(\lambda ^{-1/2})$ for
large $\lambda $ implies $g'(\lambda )=O(\lambda ^{-3/2})$, which leads to the
following moment conditions,

\be \label{mom1}
\int_{\cup G_j}\frac{\rho'(\lambda )-\alpha'(\lambda
)}{\sqrt{R_{2N+1}^+(\lambda)}} \lambda ^kd\lambda +\int_{(0,E)\setminus\cup
G_j} \frac{-i\tau'(\lambda)}{\sqrt{R_{2N+1}^+(\lambda)}}\lambda ^kd\lambda
=0\, ,
\ee
$$
k=0,\dots,N \, .
$$

A second set of conditions is obtained by
integrating $g'$ around each $G_j$ and using (\ref{ebands0}):
In case A we
obtain
\be \label{mom2}
\int_{G_j}(g'_+(\lambda )-g'_-({\lambda }))d\lambda
=-2i(\tau(r_{2j+1})-\tau(r_{2j})), \ \ j=1,\dots N \, .
\ee

In case B, $\tau(r_{2N})$ and $\tau(r_{2N+1})$ must be replaced by zero.

Conditions (\ref{mom1}) and (\ref{mom2}) represent  a system of
$(N+1)+N=2N+1$ independent equations for the $2N+1$ unknowns (the branch
points $r_1,\dots, r_{2N+1}$ of the Riemann surface (\ref{R})).
Conversely suppose that for given $x$, $t$ and some
$N$, the quantities $r_1,\dots, r_{2N+1}$ satisfy conditions
(\ref{mom1}) and (\ref{mom2}), giving rise to an explicit
expression (\ref{g'})
for $g'$, and hence for $g$ by integration.
Suppose further that the function
$g$ so constructed, also satisfies the inequalities (\ref{igaps0})
-- (\ref{B0}). Then $g$ is the desired solution of the scalar RH problem.

Deift, Venakides and Zhou show that the parameters $r_j$ represent
the ``semiclassical spectrum" of the problem, i.e. they are
the branch points of the spectral Riemann surface defining the
local finite-gap solution of the KdV.
In Sec.4 we shall directly identify the moment conditions (\ref{mom1}),
(\ref{mom2}) with the local solution of the KdV-Whitham system.

Now we present a number of important relations which will
be very useful for the identification.

We observe from the relations (\ref{egaps0}) -- (\ref{>00}),
that $g(\lambda)$
also satisfies a RH problem. Solving this RH problem in
exactly the same way as
the problem for $g'$  we obtain similarly to (\ref{g'})

\be \label{g}
g(\lambda)=\sqrt{R_{2N+1}(\lambda)}\sum_{j=0,\dots,N} \left( \int_{G_j}
\frac{2\rho_+(\mu)-2\alpha(\mu)-\Omega_j}
{\sqrt{R_{2N+1}^+(\mu)}(\mu-\lambda)}\frac{d\mu}{2\pi i}+
\int_{(0,E)\setminus\cup G_j}
\frac{-2i\tau(\mu)}{\sqrt{R_{2N+1}^+(\mu)}(\mu-\lambda)}\frac{d\mu}{2\pi i}
\right) \, .
\ee

The expression for the constant of integration $\Omega _j$ is

\be \label{omega1}
\Omega_j=
-2x\oint_{a_\infty}\lambda^{1/2}\psi_j
-8t\oint_{a_\infty}\lambda^{3/2}\psi_j+4\int_{\cup G_k}
\rho_+\psi_j-4\int_{(0,E)\setminus\cup G_k}i\tau\psi_j \, .
\ee

$$
\equiv x\Omega_{k1}+t\Omega_{k2}+\Omega_{k3} \, \ \ j=1, \dots, N \, , \
\Omega_0=0 \, .
$$
We recall that
\be \label{AB}
E=1  \ \ \hbox{ (case A)} \ \ or \ \  E=r_{2N+1} \ \ \hbox{(case B)}
\ee
The basis of holomorphic differentials $\psi _j$ is given
by
\be \label{psi}
\psi_j = \sum \limits _{k=0}^{N-1}c_{k,j} \frac{\lambda^{k}}{\sqrt { R _{2N+1}
({\bf r}, \lambda)}}d\lambda \, ,
\ee
\be \label{norm}
\oint \limits_{\alpha _k}\psi _{j} = \delta _{jk}\, ,\ \ k,j = 1, \dots, N.
\ee
The contours $\alpha_k\, ( k=0,1, \dots, N) $ surround
the intervals $(r_2, r_3), \dots , (r_{2j}, r_{2j+1}), \dots ,
(r_{2N}, r_{2N+1})$ clockwise;

The identities hold for $\Omega_{j1}$ and $\Omega_{j2}$:

\be \label{ores}
\Omega_{j1} = -Res_{\infty} \lambda^{1/2}\psi_j \, , \ \
\Omega_{j2} = -4Res_{\infty} \lambda^{3/2}\psi_j \, ,
\ee

\be \label{der}
\partial_{x}\Omega_j=\Omega_{j1}\, , \ \
\partial_{t}\Omega_j=\Omega_{j2}\, .
\ee

One can observe from (\ref{ores}) that in normalization (\ref{norm})
accepted, $\Omega_{j1}$ and $\Omega_{j2}$ can be identified with
the wave number $k_j$ and the frequency $\omega_j$ correspondingly
\cite{FFM},
where $j$ is the number of the phase , $j=1, \dots, N$
Really (taking into account the change of the normalization in comparison
with \cite{FFM}), we have the expansion as $\lambda \to \infty$
\be \label {dec}
\psi_j=\frac{1}{\lambda^{3/2}}(k_j+\frac{1}{4\lambda}\omega_j+...);
\ee
where
\be \label {}
k_j=c_{N-1,j}, \ \  \omega_j = 2c_{N-1,j} \sum \limits^{2N+1}_{m=1}r_m +
 4c_{N-2, j}\, .
\ee

Taking the mixed derivatives of (\ref{der}) we arrive at $N$ equations
expressing the wave number conservation laws, for each of $N$ phases.
\be \label{k}
\partial _t k_j =\partial _x \omega _j\, \ \ , j=1, \dots, N.
\ee

\section{Equations for the Branch Points in a Symmetric Form.
Identification with Local Solutions to the Whitham Equations}

Now we obtain equations for the branch points $r_j$
of the Riemann surface, equivalent to the
moment conditions (\ref{mom1}),(\ref{mom2}), in a symmetric
form.

Calculating $\partial_x\Omega_{j}$ directly from (\ref{omega1})
we obtain
\be \label{omx}
\partial_x \Omega_{j} = \Omega_{j1} + \sum \limits_{k=1}^{2N+1}
\frac{\partial \Omega_{j}}{\partial r_k}\partial_x r_k \, , \ \ j=1, \dots, N \, ,
\ee
which, together with (\ref{der}), implies

$$
\sum \limits_{k=1}^{2N+1}
\frac{\partial \Omega_{i}}{\partial r_k}\partial_x r_k =0 \,
$$
for any solution $r_k(x,t)$.

Thus,
\be \label{dom}
\partial_j\Omega_i=0\, ,
\ee
$$
\partial _j \equiv \frac{\partial}{\partial r_j}, \ \ j=1, \dots, 2N+1, \ \
i=1, \dots, N \, ,
$$
provided $\partial _x r_j\ne 0 \, , \qquad j=1, \dots, 2N+1$.

The system (\ref{dom}) is the system of $N(N+1)$ algebraic equations for
$2N+1$ variables $r_j(x,t)$.
Due to the uniqueness of the solution for the Riemann-Hilbert problem
(\ref{egaps0}) -- (\ref{>00})
it has to be equivalent to the moment conditions (\ref{mom1}),(\ref{mom2}).
To show that it is enough to prove the consistency of the
system (\ref{dom}). In other words
one has to show that all $N$ closed systems
for $r_j (j=1,...,2N+1)$ forming the overdetermined system (\ref{dom}) are
equivalent which is to say  that (\ref{dom}), in fact, does not depend on
the index $i$.

With this aim in view we present the following lemma.

\noindent {\bf \underline {Lemma} (Consistency) }

The overdetermined system
$\partial_j\Omega_i=0\, , \ j=1, \dots, 2N+1 \,
, \ i=1, \dots, N $ , where $\Omega_i$ is defined by (\ref{omega1})
is consistent and equivalent to the  symmetric system
of $2N+1$ algebraic equations with respect to $2N+1$ variables $r_j$:

\be \label{Lem}
\oint\limits_{a_{\infty}} \{x\lambda^{1/2}
+4t\lambda^{3/2} \}\Lambda_j = \oint \limits_{\cup \alpha_k}
\rho_+(\lambda)\Lambda_j - i \oint\limits_{\cup \beta_n
\setminus \beta_E} \tau(\lambda) \Lambda_j\, \, .
\ee

The contours $\alpha_k, \ (k=0, \dots, N)$ as it was mentioned above surround the
intervals $G_k$ clockwise.
The contours $\beta_n,  \  (n=1, \dots, N+1) $ surround
the intervals $B_n$ clockwise, and the contour $\beta_E$
surrounds the interval
$( E, 1)$, where $E=1$ (case A) or $E=r_{2N+1}$ (case B).

The differential $\Lambda_j, \  \ j=1, \dots, 2N+1$ is defined by
\be \label{Lambda}
\Lambda _j= \frac{\partial_j \psi_i}{\partial_j k_i}=
\frac {\lambda^N+ \sum
\limits^{N}_{k=1}\lambda^{N-k} p_{k,j}}{(\lambda-r_j)
\sqrt{R_{2N+1}}}d\lambda\, ,
\ee
and the coefficients $p_{k,j}$ can be  found unambigously from the
normalization conditions
\be \label{normL}
\oint \limits_{\alpha_m}\Lambda _j=0 \, , \ \ m=1, \dots , N\, ,
\ee

\vspace{0.5cm}
The proof of the lemma can be found in Appendix 1.

\vspace{0.5cm}

\noindent{\bf \underline{Theorem}  (Identification)}
The system of algebraic equations (\ref{Lem}) implicitely
defining positions of the branch points $r_j(x,t)$ of the hyperelliptic
Riemann surface in the zero dispersion limit of the RH
problem for the KdV equation satisfies
the $N$-phase averaged Whitham-KdV system \cite{FFM}, \cite{LL}.

\be \label{Wh}
\partial_t r_j=V_j({\bf r})\partial_x r_j\, , \ \  j=1,\dots, 2N+1\, ,
\ee
where $V_j({\bf r})$ are computed as certain combinations
of complete hyperelliptic integrals  \cite{FFM}.

\noindent{\bf \underline {Proof}}

To identify algebraic system (\ref{Lem})  with the solution
of the Whitham equations (\ref{Wh}) we make use of the Tsarev result
\cite{Ts}, \cite{DN}:

If $W_j(r_1,\dots ,r_{2N+1})$ is a solution of the linear overdetermined
consistent system
\be \label{Ts}
\frac{\partial _i W _j}{W_i - W_j} = \frac{\partial_i V _j}{V_i - V_j} \ ;
\ee

$$
i \ne j ; \ i,j = 1, \dots , 2N+1 \, ,
$$
then the system of algebraic equations (generalized hodograph
transform)
\be
x + V_j t = W_j  \label{hod}
\ee
gives implicitely the smooth solution $r_j(x,t)$
to the Whitham system (\ref{Wh}) provided $\partial _x r_j \ne 0$.

First, we observe that the following identities hold

\be \label{id1}
-\frac{1}{2\pi i}\oint \limits_{a_{\infty}} \lambda ^{1/2}\Lambda _j =1\,
\qquad \hbox{for all $j$} \, ,
\ee

\be \label{id2}
\frac{2}{\pi i}\oint \limits_{a_{\infty}} \lambda ^{3/2}\Lambda _j =
V_j \, .
\ee

The first equality immediately follows from the definition (\ref{Lambda}).
The second one requires a bit more detailed consideration.
We make use of the fact that the system (\ref{Wh}) implies
existence of the wave number conservation law (\ref{k}). Then, introducing
the Riemann invariants $r_j$ into the equation (\ref{k})
explicitely one easily gets the representation for the characteristic
speeds of the Whitham system (\ref{Wh}) (see \cite{Kud}, \cite{GKE} for the
case $N=1$ and \cite{El} for arbitrary $N$).  \be \label{V}
V_i=\frac{\partial_i\omega_j}{\partial_ik_j}\,  \ \ \hbox{for any $j$}\, .
\ee
The expression (\ref{V}) can be interpreted as a  generalization
of the group velocity notion to the case of nonlinear waves \cite{El}.
It should be
noted that the relationships analogous to (\ref{V}) arise in the classical
theory of hyperbolic systems as the compatibility condition providing the
existence of an additional  conservation law \cite{Lax}.  Using (\ref{ores}),
(\ref{dec}), (\ref{Lambda}) we arrive at
\be \label{Vi} V_i= \frac{2}{\pi
i}\frac{\partial_{i}}{\partial_i k_j}\oint \limits_{a_{\infty} }\psi_j\lambda
^{3/2} d\lambda= \frac{2}{\pi i}\oint \limits_{a_{\infty}} \lambda
^{3/2}\Lambda _j = -4p_{1,i} \, .
\ee
The solution  (\ref{Lem}) then takes the
form (\ref{hod}) provided
\be \label{sol}
W_i= -\frac{1}{2\pi i}\{ \oint\limits_{\cup\alpha_k}
\rho_+(\lambda)\Lambda_i-
i\oint \limits_{\cup \beta_n \setminus\beta_E}\tau(\lambda)\Lambda_i\}\, ,
\ee
where the functions $\rho_+(\lambda) \, , \tau(\lambda)$ are supposed to be
analytic.

We have to prove, therefore, that  (\ref{sol})
does solve the linear system (\ref{Ts}) which implies proving
that $\Lambda_j$ does solve this system at any $\lambda$.
With this aim in view we consider the following combinations occuring
in the left-hand part of (\ref{Ts})
\be \label{comb1}
\partial_i\Lambda_j = \partial_ip_{1,j}
\frac {\lambda^N+ \sum \limits^{N}_{k=1}\lambda^{N-k}
a_{k,i}}{(\lambda-r_i)(\lambda-r_j)\sqrt{R_{2N+1}}}d\lambda
\, ,
\ee
where all $a_{k,i}$ are uniquiely defined by the
conditions following from the normalization (\ref{normL})
\be \label{norm1}
\oint \limits_{\alpha_m}\partial_i \Lambda _j=0 \, , \ \ m=1, \dots , N.
\ee
Another relevant combination is
\be \label{comb2}
\Lambda _i-\Lambda _j=(p_{1,i}-p_{1,j})
\frac {\lambda^N+ \sum \limits^{N}_{k=1}\lambda^{N-k}
b_{k,i}}{(\lambda-r_i)(\lambda-r_j)\sqrt{R_{2N+1}}}d\lambda \, ,
\ee
where, again, the coefficients $b_{k,j}$ are given by the
normalization
\be \label{norm2}
\oint \limits_{\alpha_m}(\Lambda _i-\Lambda _j)=0 \, , \ \ i\ne j\, ,
\ee
that implies $b_{k,i}=a_{k,i}$.
Then
\be \label{eq}
\frac{\partial_i\Lambda_j}{\Lambda _i-\Lambda _j}=
\frac{\partial_ip_{1,j}}{p_{1,i}-p_{1,j}} \, ,
\ee
Recalling that $p_{1,j}=-1/4V_j$ (see (\ref{Vi})) we prove the
identification theorem.

Q.E.D.

We also present the equivalent form of the solution
(\ref{Lem}) parametrized by the
phase of the reflection coefficient from the left
$\rho_-$ (cf. (\ref{rho+})),
\be \label{rho-}
\rho_-(\lambda)=\int \limits_{-\infty}^{x^{-}}[\lambda^{1/2}-
(\lambda-u_0(x))^{1/2}]dx - \lambda^{1/2}x^{-}\,
\ee
It can be shown that the following identity holds
(for details see Appendix 2)

\be \label{iden}
\oint \limits _{\cup\alpha_k} \{\rho_-(\lambda)+\rho_+(\lambda)\}\Lambda_j-
i\oint \limits _{\cup \beta_n}\tau(\lambda)\Lambda_j=0\, .
\ee
Then the solution (\ref{Lem}) takes the form
\be \label{Lem1}
\oint\limits_{a_{\infty}} \{x\lambda^{1/2}
+4t\lambda^{3/2}\}\Lambda_j = -\oint \limits_{\cup \alpha_k}
\rho_-(\lambda)\Lambda_j + i \oint\limits_{\beta_E}\tau(\lambda)
\Lambda_j\, ,
\ee

In particular, in the case A we have especially simple representation
(we note that $\cup\alpha_k=a_{\infty}$)

\be \label{A1}
\oint\limits_{a_{\infty}} \{x\lambda^{1/2}
+4t\lambda^{3/2}+
\rho_-(\lambda)\}\Lambda_j=0\, , \  \ j=1,\dots,2N+1 \, .
\ee
This form of the solution coincides with the one obtained in \cite{El}.

The differential $\Lambda_j$ for the first time was introduced by El
\cite{El} as a
generating differential of the Whitham hierarchy and can be regarded
as a nonlinear analog of the Green function for the modulation
equations.
Really,
as $\Lambda_j$ depends on a free parameter $\lambda$ and
satisfies identically the Tsarev equations (\ref{Ts}), we have
proved automatically that it is a fundamental solution to the
Tsarev  system .
Expanding (\ref{Lambda}) in powers of $1/\lambda$ as $\lambda \to \infty$
we obtain the homogeneous solutions to Tsarev's equations. Those of
them with the odd indices of homegeneity $n=3,5,7\dots$
are the characteristic speeds of the $N$-gap averaged $n$-th KdV in
the hierarchy \cite{Kr88}. Therefore, $\Lambda_j$ is the generating
differential for the averaged KdV hierarchy. The effective formulae for the
characteristic speeds of the hierarchy in terms
of the generating differential are (up to a norming constant):
\be \label{Wn}
W^{(n)}_j({\bf r})d\lambda = -Res_{\infty}\lambda^{n-3/2}\Lambda_j\, ,
\ee
Corresponding solutions $r_j(x,t)$  given by the system
\be \label{Wss}
\oint\limits_{a_{\infty}} \{x\lambda^{1/2}
+4t\lambda^{3/2}+
c \lambda^{n-3/2}\}\Lambda_j=0\, , \  \ j=1,\dots,2N+1 \, .
\ee
are self-similar \cite{Kr88}
\be \label{ssr}
r_j(x,t)=t^\gamma l_j (\frac{x}{t^{\gamma+1}})\, , \ \
\gamma =\frac{1}{n-1}\, , \  n=3,5,\dots,
\ee
and solve the initial value problem
\be \label{ss0}
x=[2^n\frac{(2n-1)!!}{n!}cu^n]_{t=0}\, .
\ee
Taking $n=3$ we arrive at the Potemin solution \cite{P} (see also \cite{DN})
describing the universal regime of the formation
of collisionless shock in the vicinity of the break point \cite{GP},
\cite{DN}, \cite{GKE}.  This solution was also obtained by Wright
\cite{Wr} applying
the Lax-Levermore methods \cite{LL}.  We also note that if
\be \label{kr}
\rho_-(\lambda)=-\sum\limits_{k=2}^{2N+1}c_k\lambda^{k+1/2} \, ,
\ee
then the formula (\ref{A1}) gives the realization of the general
Krichever prescription for constructing the algebraic-geometrical
solutions to the Whitham equations \cite{Kr88}.

\section{Functionals of the Peierls-Fr\"ohlich type
as the Potentials for the Local Solutions to the Whitham
Equations}

We point out the important relationship between the differential
$\Lambda _j$ and the standard meromorphic differential (quasimomentum)
\be \label{dp}
dp= \frac {\lambda^N+ \sum \limits^{N}_{k=1}\lambda^{N-k}
q_k}{\sqrt{R_{2N+1}}}d\lambda
\ee
normalized by

\be \label{normp}
\oint \limits_{\alpha_m}dp=0\, , \ \ m=1, \dots, N \, .
\ee
As can be easily seen,
\be \label{lp}
\Lambda _j=\frac{\partial_jdp}{\partial_j q_1 +\frac{1}{2}}\,
\ee
In particular, for $N=0$:
\be \label{l0}
\Lambda=2\partial _r dp=\frac{d\lambda}{(\lambda-r)^{3/2}}
\ee
It is well known (see, for instance, \cite{DN}) that $\int dp$ is
the generating function for the averaged Kruskal integrals $Q_k$:
\be \label{P}
P=\int dp=2\sqrt\lambda+\sum\limits_{k=0}^{\infty}\frac{2Q_k}{(2\sqrt\lambda)
^{2k+1}}\, .
\ee
Expanding (\ref{dp}) near infinity and comparing with (\ref{P}) we find
that
\be \label{qQ}
q_1=Q_0-\frac{1}{2}\sum\limits^{2N+1}_{j=1}r_j\, ,
\ee
where
\be
Q_0=\bar u =\lim \limits_{L\to\infty}\frac{1}{2L}\int\limits_{-L}^{L}
u(x,{\bf r})dx
\ee
Then it follows from (\ref{lp}) that
\be \label{Lp}
\Lambda _j=\frac{\partial_jdp}{\partial_j \bar u}\,
\ee
provided $\partial_j \bar u \ne 0$.
Substituting the representation (\ref{Lp}) into the solution (\ref{Lem})
one arrives at the potential form of the local solution to the Whitham
equations

\be \label{Min}
\partial_j F_N({\bf r};x,t)=0 \, , \ \ j=1,\dots,2N+1 \, ,
\ee
where the potential

\be \label{FN}
F_N({\bf r};x,t) = \Phi [u_0(x);x,t,N]= \frac{1}{2\pi i}
[\oint\limits_{a_{\infty}} \{x\lambda^{1/2}
+4t\lambda^{3/2}\}dp +2\int \limits_{\cup G_k}\rho_-(\lambda)dp
-2i\int \limits_E ^1 \tau(\lambda)dp] \, ,
\ee
In the case A (E=1) we observe that $F_N$ represents the
functional of the Peierls--Fr\"ohlich type \cite{Bel}, \cite{Kr},
\cite{BDK}:

\be \label{FNA}
F_N({\bf r};x,t) =  \frac{1}{2\pi i}
[\oint\limits_{a_{\infty}} \{x\lambda^{1/2}
+4t\lambda^{3/2}\}dp +2\int \limits_{\cup G_k}\rho_-(\lambda)dp )]\, .
\ee

The case of the Peierls--Fr\"ohlich type functional with (cf.(\ref{kr}))
\be \label{pfr}
\rho_-(\lambda)=-\sum\limits_{k=2}^{2N+1}c_k\lambda^{k+1/2}
+ 2\int \limits_{\lambda}^{\infty}\frac{g(u)}{\sqrt{u-\lambda}}du\, ,
\ee
where $g(u)$ is a sufficiently small smooth function, has been
studied recently by Dubrovin \cite{D} who found that the
minimizer to (\ref{FNA}), (\ref{pfr}) gives the solution to the
Cauchy problem for the KdV-Whitham system with the monotone initial data
\be\label{pfid}
x=[\sum\limits_{k=2}^{2N+1} \frac{(2k+1)!!}{2^{k-1}k!}c_ku^k +g(u)]_{t=0} \, .
\ee
One can suppose that the minimizer to the functional (\ref{FN})
gives rise to the solution to the Cauchy problem with
hump-like initial data.
We emphasize  also that the solution in the form
(\ref{Min}), (\ref{FN}) does not require
analyticity from the initial data.

\section{Global Solutions to the Whitham System}
\subsection{General Formulation}
As it was shown in Sec.2 which gives an account of the results
of \cite{DVZ},
the steepest descent for the RH problem yeilds the global solution
to the Whitham equations bypassing the procedure of integration
of the modulational
system itself. The obtained solution, however, is restricted by the
requirements of decaying at infinity and of analyticity imposing upon
the initial data. In addition, the consideration in \cite{DVZ} concerns
only with solitonless (pure reflective) initial data. On the other
hand, the Whitham equations themselves admit more general formulation
of the problem cancelling these restrictions.

Our idea is to construct the global solutions to the general Cauchy
problem for the Whitham equations by applying {\it the results}
of the RH problem approach described above to a more
general class of functions $\rho(\lambda)$, $\tau(\lambda)$
(and therefore to a more general class of initial data) appearing in
the  solution (\ref{Lem}). More in detail, the functions
$\rho(\lambda)$, $\tau(\lambda)$ can be multivalued (even infinite-valued)
with different number of branches for different $\lambda$.
Actually, as we will show, such type of behavior for $\rho$ and $\tau$
corresponds to a multihump (or infinite-hump nondecaying) initial data.
Also, the resulting formulation of the RH problem does not require
any analyticity from the functions $\rho(\lambda)$ and $\tau(\lambda)$
which cancels requirement of analyticity for the initial data and
is consistent with the hyperbolic nature of the zero-dispersion KdV
limit \cite{Lev}.

First, we define following Dubrovin \cite{D} the Whitham system
as a sequence of the modulation systems (\ref{Wh}) defined for
$N=1,2,\dots$ For $N=0$ this coincides with the Riemann
wave equation
\be \label{RW}
\partial_t r - 6r \partial_x r =0 \, ,
\ee
with the initial data $r(x,0)=u_0(x)$ are given.
Solutions of the Whitham equations (\ref{Wh}) for a given $N$
typically exist only within certain domains of the $(x,t)$ plane.
The main problem of the theory of the Whitham equations is to
glue together these solutions in order to produce a $C^1$-smooth
multivalued function of $x$ that also depends $C^1$-smoothly
on the parameters $r_1, \dots,r_{2N+1}$.

Thus, to produce the global solution to the Whitham system with the aid
of the obtained local solutions (\ref{Lem}) one should:

i) determine the right genus at every point $x_0, t_0$.

ii) provide $C^1$-smooth matching of the solutions ({\ref{Lem}}) for
different genera on the phase transition boundaries \cite{DN}, \cite{D}.

The properties of the phase transitions can be investigated inside
the local Whitham theory.

\subsection{Phase Transitions}

We study what happens to the solution (\ref{Lem}) when
one of the Riemann invariants $r_{2j}$ coalesces either with
$r_{2j-1}$ or with $r_{2j+1}$. It follows from the solution
(\ref{Lem}) that its phase transition properties are
completely determined by the properties of the differential
$\Lambda_j$ near the double points $r_k=r_{k+1}$ and
can be investigated directly using the representation (\ref{Lambda})
for $\Lambda_j$.  It is more convenient, however, to use the relationship
(\ref{Lp}) between $\Lambda_j$ and the meromorphic differential
$dp$ properties of which are known well.

Near the double points, the differential $dp$ as well as the
coefficients $Q_k$ in the decomposition (\ref{P}) have the
following asymptotics \cite{D}:

$1)\ \ r_{2j+1}-r_{2j}\to 0$ (small gap)
\be \label{fg}
f_N(r_1,\dots,r_{2N+1})= f_{N-1}(r_1,\dots,\hat r_{2j},\hat r_{2j+1},
\dots, r_{2N+1})+ \nu^2f_{N,j}^1(r_1,\dots,\hat r_{2j},\hat r_{2j+1},
\dots; \beta,\nu) + o(\nu^2)\,.
\ee
Here
\be \label{bn}
\beta=\frac{r_{2j}+r_{2j+1}}{2}\, , \ \ \ \nu =\frac{r_{2j+1}-r_{2j}}{2}\, .
\ee

Here and below the hat means that the correspondent coordinate is
omitted.

$2)\ \ r_{2j}-r_{2j-1}\to 0 $ (small band)
\be \label{fb}
f_N(r_1,\dots,r_{2N+1})= f_{N-1}(r_1,\dots,\hat r_{2j-1},\hat r_{2j},
\dots, r_{2N+1})+ \delta f_{N,j}^2(r_1,\dots,\hat r_{2j-1},\hat r_{2j},
\dots; \eta,\delta) + o(\delta)\,.
\ee
Here
\be \label{vd}
\eta=\frac{r_{2j-1}+r_{2j}}{2}\, , \ \ \ \delta =[\log \frac{4}{(r_{2j}-r_{2j-1})^2}
]^{-1}\, .
\ee

One can see that in both cases, in the limit, the double
points drop out of the function $f_N$ and it turns into its
analog for the $N-1$ genus. This fact follows from the normalization
for the meromorphic differential
\be \label{np}
\int\limits_{r_{2j}}^{r_{2j+1}}dp=0\, , \qquad j=1, \dots, N \, ,
\ee
which implies that the polynomial $\lambda^N + \sum\limits_{k=1}^{N}
q_k\lambda^{N-k}$ in the numerator of $dp$ has exactly one
zero in each gap. Then, if one shrinks either gap or band,
this zero inevitably coincides with the double point in the denominator.

Certainly, all the averaged Kruskal integrals $Q_k(r_1,\dots,r_{2N+1})$
(see (\ref{P})) have the same properties near the double points.
Then using the
representation (\ref{Lp}) for the generating differential we arrive at the
following asymptotics by differentiating (\ref{fg}) and (\ref{fb}).

$1) \ \ r_{2j+1}-r_{2j}\to 0$
\be \label{Lg}
\Lambda^{[N]}_k (r_1,\dots,r_{2N+1};\lambda)=\Lambda_k^{[N-1]}
(r_1,\dots,\hat r_{2j},\hat r_{2j+1},
\dots, r_{2N+1};\lambda)+ O(\nu^2) \, ,
\ee
$$
\hbox{if} \ \ k\ne2j,2j+1\, ,
$$
and
\be \label{Lg+}
\Lambda^{[N]}_k (r_1,\dots,r_{2N+1};\lambda)=
\Lambda_{2j+}^{[N]} + O(\nu) \, ,
\ee
$$
\hbox{if} \ \ k=2j \ \ \hbox{or} \ \ k=2j+1\, ,
$$
where $\Lambda_{2j+}^{[N]}$ is the limiting value of the differential
$\Lambda_k^{[N]}$ which follows from (\ref{Lambda}), (\ref{normL})
when one pinches the $j$-th gap :
\be
\label{Lk+} \Lambda_{2j+}^{[N]} \equiv \Lambda_{k}^{[N]} (r_1,\dots, r_{2j-1},
r_{2j},r_{2j}, r_{2j+2},\dots, r_{2N+1};\lambda)
=\frac{\lambda^{N}+\dots}{(\lambda-r_{2j})^2\sqrt{R'_{2N-1}}}d\lambda \, ,
\ee
$$
R'_{2N-1}= (\lambda-r_1)(\lambda-r_2)\dots(\lambda-r_{2j-1})(\lambda-r_{2j+2})
\dots(\lambda-r_{2N+1}),
$$
and $\Lambda_{2j+}^{[N]}$ has a double pole at $\lambda=r_{2j}$.

$2) \ \  r_{2j}-r_{2j-1}\to 0$
\be \label{Lb}
\Lambda^{[N]}_k (r_1,\dots,r_{2N+1};\lambda)=\Lambda_k^{[N-1]}
(r_1,\dots,\hat r_{2j-1},\hat r_{2j},
\dots, r_{2N+1};\lambda)+ O(\delta) \, ,
\ee
$$
\hbox{if} \ \ k\ne2j-1,2j\, ,
$$
and
\be \label{Lb-}
\Lambda^{[N]}_k (r_1,\dots,r_{2N+1};\lambda)=
\Lambda_{2j-}^{[N]} + O(\nu/\delta) \, ,
\ee
$$
\hbox{if} \ \ k=2j-1 \ \ \hbox{or}\ \  k=2j\, ,
$$
where the $\Lambda^{[N]}_{2j-}$ is the limiting value of the differential
$\Lambda_k^{[N]}$
when one pinches the $j$-th band :
 \be \label{Lk-} \Lambda_{2j-}^{[N]} \equiv
\Lambda_{k}^{[N]} (r_1,\dots, r_{2j-2},r_{2j},r_{2j}, r_{2j+1},\dots, r_{2N+1};\lambda)
=\frac{\lambda^{N-1}+\dots}{(\lambda-r_{2j})\sqrt{R'_{2N-1}}}
d \lambda \, ,
\ee
$$
R'_{2N-1}= (\lambda-r_1)(\lambda-r_2)\dots(\lambda-r_{2j-2})(\lambda-r_{2j+1})
\dots(\lambda-r_{2N+1}),
$$
and $\Lambda^{[N]}_{2j-}$ has a single pole at $\lambda=r_{2j}$ (cancellation of one
pole occurs due to the zero in the vanishing $j$-th band ).

One can see the substantial difference in the limiting behavior
of the differential $\Lambda_{2j} ^{[N]}$ depending whether one shrinks
the $j$-th gap (\ref{Lk+}) or the $j$-th band (\ref{Lk-}).

The asymptotics (\ref{Lg}) and (\ref{Lb}) provide
natural $C^1$ - smooth matching of the solution (\ref{Lem})
for different genera on the phase transition boundaries.
The limiting values in (\ref{Lg+}) and (\ref{Lb-})
determine the motion of those boundaries.
Namely, the boundaries $x_{j\pm}$ separating  $N$-phase and $N-1$-phase
regions correspond
to closing either the $j$-th gap $(+)$ (linear wave degeneration) or
the $j$-th band $(-)$
(soliton degeneration) and obey the ODE's :
\be \label {b+-}
\frac{dx_{j\pm}}{dt}= -4 Res_{\infty}\lambda^{3/2}\Lambda_{2j\pm}^{[N]}
\ee

One should also check what happens to the solutions
(\ref{Lem})  under the phase transition $(N=1) \to (N=0)$.
As $N=0$ we have $\Lambda_j^{[0]}=(\lambda-r)^{-3/2}d\lambda$ (see
(\ref{l0})) and, as a result, we arrive at two different
equations corresponding to the cases A and B respectively.
In the case A we have

\be \label{A00}
\hbox {A:} \qquad \ \  \ x-\frac{2t}{\pi i}\oint \limits_{a_{\infty}}
\frac{\lambda^{3/2}}{(\lambda-r)^{3/2}}d\lambda =
\frac{2}{\pi }\frac{d}{dr}\int \limits_{0}^{r}
\frac{\rho_-(\lambda)}{\sqrt{r-\lambda}}d\lambda\, ,
\ee
which is  the solution of the Riemann wave equation (\ref{RW})
with the initial data in the form of
the increasing part of the hump $u_0(x)$:
\be \label{A01} x+6rt=x^-(r)\, .  \ee
For the
case $B$ we have an analogous result
for the decreasing part of the initial hump $u_0$ :
\be \label{B01}
\hbox{B:} \qquad \ \ x+6rt=x^-(r)- \frac{2}{\pi }\frac{d}{dr}\int
\limits_{r}^{1} \frac{\tau(\lambda)}{\sqrt{\lambda-r}}d\lambda = x^+(r)\, .
\ee

It can be easily seen also that the decompositions of $\Lambda_j^{[1]}$
near $\Lambda_j^{[0]}$ have the form (\ref{Lg}), (\ref{Lb}) that provides
$C^1$ - smoothness of the transition $N=1 \to N=0$. This transition
was investigated in detail by Avilov, Krichever and Novikov \cite{AN}
(see also \cite{DN}, \cite{D}).

It is clear now that the cases A and B in the solution (\ref{Lem})
describe different parts of the global solution of the
Whitham equations corresponding to the
monotonic branches of the initial profile $u_0(x)$.
We also note that  formulas (\ref{A00}), (\ref{B01}) give the solution
to the Riemann equation in terms of the KdV scattering data and were
obtained in this form by Geogjaev in \cite{G}.

\subsection{Inequalities: Determination of Genus}

As it has been said before, the necessary part of constructing the
solution to the Cauchy problem is determination of the genus
of the Riemann surface at every point $x,t$. In fact,
this problem cannot be resolved locally in the frame of the obtained solution
(\ref{Lem}) for the Whitham system. This point of view has not been
established clearly in the literature so we will describe it
more in detail.

The genus (the number of oscillating phases) $N$ is the "external"
characteristics with respect to the local solutions
to the Whitham system.  To illustrate that more
clearly, we present a simple example of ambiguity in the local
determination of genus.

In the Figure 1, there is a typical solution of one-phase
averaged Whitham
equations for the initial data with the only break point
(for example, the cubic breaking in the Gurevich -- Pitaevskii problem
\cite{GP})
depicted by a solid line in the region $(x_-,x_+)$ while the
three-valued solution to the Riemann wave equation (\ref{RW}) is
depicted by a broken line in the region $(x_1, x_2)$,
$x_1<x_-<x_2<x_+$. Outside of those
intervals both solutions coincide.

One can see that in the vicinity of any point of an open interval
$(x_2,x_+)$ both cases $g=0$ and $g=1$ can be applied without any violations
of the existence for the local solution. The missing part of the
information which enables to distinguish the unique solution in this interval
is the break time $t_{crit}$ (if $t>t_{crit}$  then $N=1$ otherwise $N=0$)
but this information
is relevant to the global properties of the solution for
 case $N=1$.
In other words, generically to make the right decision about the genus at
$t=t_0$ one should follow the
solution all time until $t_0$ and change genus after passing the critical
points .
The rigorous global predictions of the genus
have been made so far only in a few simple cases.

In the case of analytic initial data with the only break point
there are results by Tian \cite{T1}, \cite{T2} that the solution beyond the
break time globally belongs to $N=1$.

Another result in this area is due to Grava \cite{Gr} and it reads that
the maximal genus in the Cauchy problem with polynomial initial data
does not exeed the degree of the polynomial and $N\to1$ asymptotically
as $t\to\infty$ (the latter result also can be found in \cite{El}).

But even in the simplest case of the polynomial data the determination
of genus at a fixed point $x_0,t_0$ requires knowledge of the behaviour
of the solution in all earlier times.

The RH problem approach makes it possible to determine
the right genus of the problem locally with the aid of the inequalities
(\ref{ibands0}) -- (\ref{>00}) for the complex phase function $g(\lambda)$.
These
original inequalities were derived for the decaying solitonless
initial data in \cite{DVZ}, \cite{V}.  It is clear, however, that if
one assumes the finite speed of propagation (hyperbolicity) in the zero
dispersion limit (which has been rigorously  proved for the case of decaying
initial potential \cite{LL}, \cite{V85}, \cite{Lev}) then one can consider an
arbitrary initial profile and regard these inequalities as a complementary
part to the general local solution (\ref{Lem}) for finite $t$.  Really, given the
functions $\rho_\pm(\lambda)$ and $\tau(\lambda)$ as the simple integral (Abel)
transforms (\ref{rho+}),(\ref{rho-}), (\ref{tau}) of arbitrary initial data one
can construct the function $g(\lambda)$ and its derivative $g'(\lambda)$ and
then, by trial and error, to determine the genus with the aid of inequalities.
Also, one should add the relations (\ref{A0}) and (\ref{B0}) distinguishing the
cases A and B for the problem with nonmonotone initial data. Certainly, for the
monotonically increasing data we always are in the case A.

{\bf Example: One-Hump Problem}

Following \cite{DVZ} we present now an example of a systematic procedure for
obtaining $N=N(x,t)$ for all $x$ at any given $t$.
The result of this procedure is to
construct for each $t$ the $separatrix$
$F(t)=\{ (x, r ): r = r(x,t), -\infty <x<\infty,
r = r_1, \dots, r_{2N+1}\}$ for the case of one-hump initial data
with finite number of break points (see Figures 2a and 2b).

For times $t$ less than some critical value $t_{crit} $ one takes
$N=N(x,t)=0$ and solves (\ref{RW}) for $r$; the time $t_{crit}$
corresponds precisely to the time at
which (\ref{KdV}) with $\epsilon =0$ breaks down. It turns out that
the associated function $g$ constructed as above indeed satisfies
the auxilliary inequalities (\ref{igaps0}), (\ref{ibands0}), (\ref{A0}),
(\ref{B0}) and
hence $g$ is  the desired solution. Thus, for $t<t_{crit}$, we obtain a
separatrix of the
shape of Figure 2a; moreover, we find that $F(t)={(x,r):  r=u(x,t)}
$, where $ u(x,t)$ is the solution of (\ref{KdV}) with $\epsilon =0$. For $
x>x_0(t)$, we are in case B and for $x<x_0(t)$, we are in case A.
When $t>t_{crit}$, one again sets $N=N(x,t)=0$ for $x>>1$, computes $r$ from
(\ref{RW}) and verifies once again that the side conditions are satisfied in
case B.  However, as we move $x$ to the left, we find that at least one of the
inequalities (\ref{igaps0}), (\ref{ibands0}), (\ref{A0}),
(\ref{B0}) breaks down. In Figure 2b below the
first inequality in (\ref{igaps0}), $ -\tau <(g_+-g_-)/2i $, breaks down at
$x=x_1$ and for $x<x_1$ the interval $G_0$ breaks up into two intervals $G_0$
and $G_1$. For such $x$ one solves the system (\ref{Lem})
for one of the cases A or B
to get $r_1, r_2$, and $r_3$ and verifies that indeed the associated $N$
satisfies the above inequalities.  In the scenario of
Figure 2b, as we move  $x$ further to the left, the same inequality for
$\tau $ again breaks down at some $x=x_2 (t)$, for some
$r \in G_0(x_2 (t),t) $.
Again $G_0$ splits up into two intervals yielding a total of three intervals
$G_0$, $G_1$ and  $G_2$, etc. As we continue to move towards $x_3$, the
interval $G_1$ shrinks to a point  and eventually disappears. For  $x$ between
$x_4$ and $x_3$ we are again in the two interval case and for $x<x_4$ we
return
to the one interval case $N=0$. For $x>x_0$, we are always in case B and
for $x<x_0$ we are always in case A.  The scenario of figure 2b is
representative of the generic case in which any finite number of intervals may
be open at some $x$. For more complicated, indeed pathological initial data
infinitely many folds may appear.

\section{Zero-Dispersion KdV with General Initial Data }

The basic assumption which is made in the following consideration
is the hyperbolicity of the zero dispersion limit for the KdV
with an arbitrary initial potential.
By applying this assumption one can extend the formulation of the RH
problem made in Sec.2 for the case of the solitonless alanytic
initial perturbation
to the case  of an arbitrary  initial data and
to construct the solution to the corresponding Cauchy problem for the
KdV-Whitham system.

It is clear that due to the (assumed) hyperbolicity of the problem
only a finite number of
`humps' in the initial perturbation will be involved into nonlinear
interaction at any finite time at any particular point.
Therefore, the solution to the  problem for the
Whitham equations with an arbitrary (nondecaying) initial data can be
obtained by solving the "multi-hump" (decaying) problem in the
zero dispersion limit and
then by tending the number of humps to infinity.
Certainly, the solution obtained in this way will be valid
during limited time interval (presumably for $t\ll 1/\epsilon$) until the
semiclassical spectrum (Riemann invariants of the Whitham equations)
begins to compete with the fine spectrum of the initial potential  and the
problem loses its hyperbolic character.

The main technical obstacle
to the direct extension of the formulation
(\ref{egaps0}) -- (\ref{>00})
to the case of a multi-hump initial data is the multivaluedness
of the functions
$\rho_+(\lambda)$ and $\tau(\lambda)$ (\ref{rho+}), (\ref{tau})
appearing in the RH problem. To
manage this difficulty it is convenient to first reformulate
the RH problem for one-hump initial data introducing the finite
reference point instead of the reference point at infinity
(scattering from the right).

\subsection{Reformulation of the One-Hump Problem for the
Finite Reference Point}

We start with the function

\be \label{g0}
g(\lambda, x, 0)= \int \limits_x^{\infty}(\lambda^{1/2}-
(\lambda - u_0(x'))^{1/2})dx' \, ,
\ee
which, as can be checked directly, solves the RH problem (\ref{egaps0}) --
(\ref{>00}) for $t=0, \ \ N=0$.

We introduce a new function
\be \label{H}
H(\lambda, x, x_0) =  \int \limits_x^{x_0}(\lambda^{1/2}-
(\lambda - u_0(x'))^{1/2})dx' \, ,
\ee
where $x_0$ is an arbitrary fixed reference point.
Without loss of generality we put $x_0$ at the maximum of $u_0(x)$
(see Fig.2a).

One can see that
\be \label{gH}
g_{\pm}(\lambda,x,0)=H_{\pm}-\lambda^{1/2}x_0 + \rho_+ \mp i \tau_+ \, ,
\ee
where
\be \label{tau+}
\tau_+(\lambda)= \int \limits^{x^{+}}_{x_{0}}(u_0(x)-\lambda)^{1/2} dx\, .
\ee

Then we have the relationships (at $t=0$)
\be \label{r1}
\frac{g_+ + g_-}{2} = \frac{H_+ + H_-}{2} -\lambda^{1/2}x_0 +
\rho_+ \, ,
\ee

\be \label{r2}
\frac{g_+ - g_-}{2i} = \frac{H_+ - H_-}{2i} - \tau_+ \, .
\ee

We also introduce
\be \label{tau-}
\tau_-(\lambda)=\tau_+(\lambda) - \tau(\lambda)=
\int \limits_{x_{0}}^{x^{-}}(u_0(x)-\lambda)^{1/2} dx \, .
\ee
Then the function $H(\lambda, x, x_0)$ is the solution of the
RH problem which follows directly from
(\ref{egaps0}) -- (\ref{>00}) :

In the region $x<x^-$ (see Fig.2a) which lies on the left from the hump
and contributes to the spectral band $B_1$
the correspondent relations follow from
(\ref{A0}) (case A) taking into account (\ref{r1}), (\ref{r2}):

\be \label{a}
\frac{H_+ - H_-}{2i}= \tau_- \, , \qquad
\frac{H_+' + H_-'}{2} < -\alpha_0'\, ,
\ee
where
\be \label{alpha0}
\alpha_0(\lambda, x) = \lambda^{1/2}(x-x_0)\, .
\ee
Analogously, for the region $x^-<x<x^+$ lying under the hump and
contributing
to the gap $G_0$ we have from (\ref{egaps0}), (\ref{igaps0}):

\be \label{b}
\frac{H_+' + H_-'}{2} = -\alpha_0'\, , \qquad
 \tau_-< \frac{H_+ - H_-}{2i}< \tau_+ \, .
\ee

And finally, in the region $x>x^+$ which lies on the right from the hump
and again contributes  to the band $B_1$ (but it is the case
B now) in the original RH problem (see (\ref{B0})) we have
\be \label{c}
\frac{H_+ - H_-}{2i}= \tau_+ \, , \qquad
\frac{H_+' + H_-'}{2} > -\alpha_0'\, .
\ee
Also we have outside the interval $(0,1)$ from (\ref{<00}), (\ref{>00})
\be \label{<01}
H_+ +H_- =0 \qquad  \hbox{if} \qquad \lambda<0,
\ee
\be \label{>11}
H_+ - H_- =0 \qquad \hbox{if} \qquad \lambda>1 \, .
\ee

Due to the finite speed of propagation in the zero dispersion KdV
with decaying initial data
the topology of the picture (Fig.2a) does not change under the
$t$-evolution (see Fig 2b). Finitely many folds which appear at
some finite $t$
determine the band-gap structure at any point $x$ so that
the region $x<x^-$ on the $(u,\ x)$ - plane of the initial data
will provide the bands $B_n,\  (n=1, \dots, N+1)$
($B_{N+1}$ corresponds here to the case A), the
region $x^-<x<x^+$ will provide the gaps $G_k,\ (k=0, \dots, N) $, and the
region $x>x^+$ always will correspond to the band $B_{N+1}$ in the case B.
Then one can formulate the time-dependent RH problem for the finite
$t$ by the substitution reflecting the linear temporal
evolution of the phase $\alpha(\lambda)$ (\ref{alpha}):
\be
\label{alpha0t}
\alpha_0(\lambda,x)
\rightarrow
\alpha_0(\lambda,x,t)= \lambda^{1/2}(x-x_0)+4\lambda^{3/2}t \,   \ee in the
formulas (\ref{a}) -- (\ref{>11}).

Thus, we have effectively removed the infinite reference point
from the formulation of the RH problem corresponding to the
evolution of the one-hump perturbation.

\subsection{Evolution of an Arbitrary Profile}

The strategy of investigation of the evolution for
an arbitrary initial perturbation $u_0(x)$ is essentially the same
as in the previous subsection: we formulate the RH problem
for the function $H(\lambda,x, x_0)$ (\ref{H}) at $t=0$ starting with the
fixed
reference point $x_0$ and then by {\it assuming} the finite speed
of propagation (hyperbolicity of the zero-dispersion KdV)
arrive at the
time-dependent $H$-function by the substitution (\ref{alpha0t})
for the phase $\alpha_0$ .
As was mentioned, the situation, however, is
complicated by the multivaluedness of the functions
$\rho(\lambda)$ and $\tau(\lambda)$ appearing in such a problem.

We begin with the multihump initial potential and
split up the strip $0<u<1$ in the $u, x$ - plane of the initial
data into two type of domains (see Fig.3a):

$\tau$ - domains correspond to  humps in the intitial
perturbation and contribute to the gaps in the semiclassical
spectrum.

$\rho$ - domains correspond to  wells in the initial
perturbation and contribute to the bands in the semiclassical
spetrum.

We fix the reference point $x_0$ at some maximum of the
function $u_0(x)$ and enumerate the monotonic parts
of $u_0(x)$ for $x>x_0$ labelling them by the roots of the
equation $\lambda = u_0(x)$ (to avoid unnecessary notation
complexities
we put the number of humps on the right from the point $x_0$
equal to the number of humps on the left):

\be \label{xn}
x_{-M}^-   < x_{-M+1}^+< \dots   < x_{-1}^+ < x_{-1}^-
\le x_0 \le x_1^+ <x_1^-  < \dots< x_{M-1}^- <x_{M}^+\, ,
\ee

$$
M \in {\bf N} \, ,\ \ \frac{dx_m^-}{d{\lambda}}>0
\, , \qquad \frac{dx_m^+}{d{\lambda}}<0   \, .
$$

We denote the domains of definition for each $x_m^{\pm}(\lambda)$
as $D_m^{\pm}$.
Then for each $\lambda=\lambda^*$ one can define two subsequences:
$\{x_{m_j}^-(\lambda^*)\} $ and
$\{x_{m_j}^+(\lambda^*)\} $,
($m_j =\pm 1, \pm 2, \dots, \pm M_j \, , \
j=j(\lambda)= 1,\dots ,2M-1 \, , \ x_{k_j}^+ > x_{k_j-1}^-) $
such that only those $x_k^{\pm}$ participate in $\{x_{m_j}^{\pm}\}$
for which $\lambda^* \in D_k^{\pm}$ (Fig3.b).

For each $j=j(\lambda)$ one can then define the sequences
$\{\rho_{m_j} \}$, $\{\tau_{m_j}\}$:

\be \label{secuence}
\rho_{m_j}(\lambda)=
Re \int \limits_{x_{0}}^{x^-_{m_j}}(\lambda-u_0(x'))^{1/2} dx' \, ,
\qquad \tau_{n_j}(\lambda)=
Im \int \limits_{x_{0}}^{x^+_{n_j}}(\lambda-u_0(x'))^{1/2} dx' \,
\ee

Then the function $H(\lambda,x,x_0,0)$ defined by (\ref{H}) and corresponding
now to a multi-hump analytic initial data $u_0(x)$ is related to the
original function $g(\lambda, x, 0)$ by (cf. (\ref{gH}))

\be \label{gH1}
\hbox{if} \ \ x^-_{m_j-1}<x<x^-_{m_j}\, ,\ \ m_j>0 : \qquad
g_{\pm}=H_{\pm}-\lambda^{1/2}x_0 + \rho_+ -\rho_{m_j-1} \mp i \tau_{m_j} \, ,
\ee where $\rho_+(\lambda)$ is redefined for the multi-hump perturbation in the
following way:

\be \label{rho+new}
\rho_+(\lambda)= x^+_{m_j} \lambda ^{1/2} + \int \limits^{\infty}
_{x^{+}_{m_j}}
[\lambda^{1/2}-(\lambda-u_0(x'))^{1/2}]dx' \, ,
\ee

Here one should put $\rho_0 \equiv 0$ and $x_{0}^{\pm} \equiv x_0$.
If $m_j<0$ then  the relationship (\ref{gH1}) is valid within the
interval $x^+_{m_j}<x<x^+_{m_j+1}$.

Then, at $t=0$ we have instead of (\ref{r1}), (\ref{r2}) within the
mentioned intervals covering the entire axis:
\be \label{r11}
\frac{g_+ + g_-}{2} = \frac{H_+ + H_-}{2} -\lambda^{1/2}x_0 +
\rho_+ - \rho_{m_j-1}\, ,
\ee

\be \label{r21}
\frac{g_+ - g_-}{2i} = \frac{H_+ - H_-}{2i} - \tau_{m_j} \, .
\ee

Now we can formulate the RH problem which is satisfied by the
function $H$. We substitute (\ref{r11}), (\ref{r21})
into the original RH problem (\ref{egaps0}) -- (\ref{>00})
taking into account new definition of $\rho_+(\lambda)$
(\ref{rho+new}):

In each $\tau$ - domain ($x^-_{m_j-1}<x<x^+_{m_j}$ for $m_j>0$
and $x^-_{m_j}<x<x^+_{m_j+1}$ for $m_j<0$ )
one has from (\ref{egaps0}), (\ref{igaps0})

\be \label{taukj}
\frac{H_+' + H_-'}{2} = -\alpha_0'+ \rho'_{m_j-1} \, , \qquad
\tau_{m_j-1} < \frac{H_+ - H_-}{2i}< \tau_{m_j} \, .
\ee

In each $\rho$ - domain  $x^+_{m_j} <x< x^-_{m_j}$
we have from
(\ref{ebands0}), (\ref{ibands0}), (\ref{B0})

\be \label{rhokj}
\frac{H_+ - H_-}{2i}= \tau_{m_j} \, , \qquad
-\alpha_0'+ \rho'_{m_j-1}<
\frac{H_+' + H_-'}{2} < -\alpha_0'+ \rho'_{m_j}\, .
\ee
We note that inequalities (\ref{ibands0}), (\ref{A0}), (\ref{B0})
have converted into one double inequality.
One  should also put $\tau_{0} \equiv 0$.

Outside the interval $(0,1)$ one has analogously to
(\ref{<01}), (\ref{>11}):
\be \label{<011}
H_+ +H_- =0 \qquad  \hbox{if} \qquad \lambda<0,
\ee

\be \label{>111}
H_+ - H_- =0 \qquad \hbox{if} \qquad \lambda>1 \, .
\ee

As well as in the previous subsection, the time-dependent RH
problem is obtained from (\ref{taukj}) -- (\ref{>111})
by adding the term $4\lambda^{3/2}t$ to $\alpha_0(\lambda, x)$
(see (\ref{alpha0t})).
However, to retain validity of the formulated RH problem after
this substitution one should {\it assume} hyperbolicity of the
zero-dispersion limit for the KdV with general initial data. In this
case the  toplogy of the separatrix would not change under the
evolution.

Recalling that $\tau$ - regions contribute to the gaps $G_k\, ,
(k=0, \dots, N)$ and $\rho$ - regions contribute to the bands $B_n\, , n=1,
\dots, N+1$, one immediately obtains the function $H(\lambda, x, x_0, t)$ and
its derivative (cf. (\ref{g}), (\ref{g'})).  For the sake of simplicity we
write down the result assuming that the initial data are such that each $\rho$
- domain contributes to only one band (analogously, each $\tau$ -- domain
contributes to only one gap). Consideration of more general data
would only complicate notations adding nothing to substance.

\be \label{H'}
H'(\lambda, x, x_0, t)= \sqrt{R_{2N+1}(\lambda)}\sum\limits_{m=0}^N
\left(
\int \limits_{G_{m}}\frac{2{\rho^*_{m}}'(\mu)- 2\alpha'_0(\mu, x, t)}
{\sqrt{R_{2N+1}^+}(\mu-\lambda)}\frac{d\mu}{2\pi i}+
\int \limits_{B_{m+1}}\frac{-2i{\tau^*_{m}}'(\mu)}
{\sqrt{R_{2N+1}^+}(\mu-\lambda)}\frac{d\mu}{2\pi i}\right) \, ,
\ee
\be \label{Ht}
H(\lambda, x, x_0, t)= \sqrt{R_{2N+1}(\lambda)}\sum\limits_{m=0} ^N
\left( \int \limits_{G_{m}}\frac{2\rho_{m}^*(\mu)- 2\alpha_0(\mu, x, t)-
\Omega_{m}}
{\sqrt{R_{2N+1}^+}(\mu-\lambda)}\frac{d\mu}{2\pi i}+
\int \limits_{B_{m+1}}\frac{-2i\tau^*_{m}(\mu)}
{\sqrt{R_{2N+1}^+}(\mu-\lambda)}\frac{d\mu}{2\pi i} \right) \, ,
\ee

where

\be \label{OmegaT}
\Omega_{m}=
-2(x-x_0)\oint\limits_{a_\infty} \lambda^{1/2}\psi_j
-8t \oint \limits_{a_\infty}\lambda^{3/2}\psi_j+
4\sum \limits_{m=0}^N (\int \limits_{G_{m}}
\rho_{m}^*\psi_j-i\int \limits_{
B_{m+1}}\tau^*_{m}\psi_j ) \, ,
\ee
$$ \Omega_0=0, \qquad \rho^*_0=0 $$.
Here
\be \label{subsets}
\{\rho^*_m (\lambda)\}\subset \{\rho_{k_j}(\lambda) \} \, , \qquad
\{\tau^*_m (\lambda)\}\subset \{\tau_{k_j} (\lambda)\}\, , \qquad
m=0,\dots, N \,.
\ee
The algebraic equations determining dependence of the band-gap structure
on $x$ and $t$ are obtained from the condition (\ref{dom})
$\partial_j\Omega_i=0$ and can be eventually represented
in a potential form (\ref{Min}) with the generalized Peierls -
Fr\"ohlich type functional as a potential (see Sec.3, 4)
\be \label{Min1}
\partial_j F_N({\bf r};x,t)=0 \, , \ \ j=1,\dots,2N+1 \, ,
\ee

\be \label{FN1}
F_N({\bf r};x,t) = \frac{1}{2\pi i}
[\oint\limits_{a_{\infty}} \{(x-x_0)\lambda^{1/2}
+4t\lambda^{3/2}\}dp -
2\sum \limits_{m=0}^N (\int \limits_{G_{m}}
\rho_{m}^*dp -i\int \limits_{
B_{m+1}}\tau^*_{m}dp )].
\ee

The system (\ref{Min1}), (\ref{FN1}) represents the general local solution
to the Whitham equations. Due to the properties of the meromorphic
differential $dp$ discussed in Sec.5.2 this solution provides
$C^1$ - smooth matching on the phase transition boundaries where
genus $N$ of the hyperelliptic surface changes. The right choice of
the local genus $N$ and the subsets $\{\tau^*_m\}, \ \{\rho^*_m\}$
(\ref{subsets}) is verifed by check of
the fulfilment of the inequalities (\ref{taukj}) and (\ref{rhokj}) for the
function $H(\lambda, x_0, x, t)$.

As an example of an effective choice of the subsets
$\{\tau^*_m\}, \ \{\rho^*_m\}$ one can indicate the case of
periodic initial data $u_0(x)$ where $\tau^*_m=(m+1/2)
Im \int_{x_0}^{x_0+T}(\lambda - u_0(x'))^{1/2}dx'$,
$\rho^*_m=m Re\int_{x_0}^{x_0+T}(\lambda - u_0(x'))^{1/2}dx'$,
$T$ is the period, $m \in {\bf N}$

Thus, we have bypassed the inverse scattering transform for
the zero - dispersion KdV with general initial data by assuming
hyperbolicity of the zero dispersion limit. One can see that
the  resulting construction
requires neither decaying nor analyticity for the initial data.

\vspace{0.5cm}
{\bf Aknowledgements}

The research described in this paper was made possible  by the
financial support of the
Civilian Research and Development Foundation for the Former
Soviet Union and USA (CRDF) under the grant $\#$ RM1 - 145.

G.A.E  and A.L.K thank also the Russian Foundation for Basic
Research (RFBR) for the
partial financial support under the grant $\#$ 00 - 01 - 00210.

\vspace{1cm}
\setcounter{equation}{0}
\def\theequation{A1.\arabic{equation}}

{\large \bf
Appendix 1: \  Proof of the Consistency Lemma}

Consider the function
\be \label{phi}
\Phi_i^{(j)}=\frac{\partial_i\Omega_j}{\partial_i k_j}\, .
\ee

We  rewrite the expression (\ref{omega1}) in
the form containing  contour integrals.
For analytic functions
$\rho_+(\lambda)$ and $\tau(\lambda)$ we have
\be \label{omegaj}
\Omega_j=
\oint\limits_{a_{\infty}} \{-2x\lambda^{1/2}
-8t\lambda^{3/2}\}\psi_j+ \oint\limits_{\cup \alpha_k}
2\rho_+(\lambda)\psi_j-2i\oint \limits_{\cup
\beta_n\setminus \beta_E}\tau(\lambda)\psi_j \, .
\ee
Here
we have used the fact that the functions $\rho_+(\lambda)$ (\ref{rho+}) and
$\tau(\lambda)$ (\ref{tau}) have their own different Riemann surfaces with the
branch points:
$$ 0,\,  \infty \qquad \hbox{for}\qquad  \rho_+(\lambda)$$  $$ 1, \,
\infty \qquad \hbox{for} \qquad \tau(\lambda) $$
Then we obtain from (\ref{phi})

\be \label{phij}
\Phi_i^{(j)}= -2
\oint\limits_{a_{\infty}} \{x\lambda^{1/2}
+4t\lambda^{3/2}\}\frac{\partial_{i}\psi_j}{\partial_i k_j} +
2\oint\limits_{\cup \alpha_k}
\rho_+(\lambda)\frac{\partial_{i}\psi_j}{\partial_i k_j}-
2i\oint \limits_{\cup \beta_n
\setminus \beta_E}\tau(\lambda)\frac{\partial_{i}\psi_j}{\partial_i k_j}\, .
\ee
We consider the differential
\be \label{lj}
\Lambda _i^{(j)}= \frac{\partial_{i}\psi_j}{\partial_i k_j}\, .
\ee
Substituting (\ref{psi}) into (\ref{lj}) we immediately obtain

\be \label{lj1}
\Lambda _i^{(j)}= \frac {\lambda^N+ \sum \limits^{N}_{k=1}\lambda^{N-k}
p^{(j)}_{k,i}}{(\lambda-r_i)\sqrt{R_{2N+1}}}d\lambda
\ee
We integrate $\Lambda _i^{(j)}$ over the $\alpha$-cycles. Then, taking into
account (\ref{norm}), (\ref{lj}) we get
\be \label{norml}
\oint \limits_{\alpha_m}\Lambda _i^{(j)}=0,
\ee
$$
m,j=1,\dots,N\, , \ i=1,\dots,2N+1
$$
The normalization (\ref{norml}) uniquely defines the coefficients $p_{k,i}$
independently on $j$, i.e.
 \be \label{equiv}
 \Lambda _i^{(1)}= \Lambda _i^{(2)}= \dots = \Lambda _i^{(N)}\equiv
 \Lambda _i \, .
 \ee
 Therefore, the differential (\ref{lj}) does not depend on the index $j$,
 and, according to (\ref{phij}), the function $\Phi_i^{(j)}$ does not
 depend on $j$ as well. That means that the system (\ref{dom}) is cosistent
and takes the form
\be \label{solA}
\oint\limits_{a_{\infty}} \{x\lambda^{1/2}
+4t\lambda^{3/2} \}\Lambda_j = \oint \limits_{\cup \alpha_k}
\rho_+(\lambda)\Lambda_j -i \oint \limits_{\cup \beta_n \setminus \beta_E}
\tau(\lambda)\Lambda_j \, ,
\ee
provided $\partial_i k_j \ne 0$.

Q.E.D.

\vspace{1cm}
\setcounter{equation}{0}
\def\theequation{A2.\arabic{equation}}

{\large \bf
Appendix 2: \  Derivation of the Identity (\ref{iden})}

We introduce the analytic function
\be \label{f}
f(\lambda) = \int \limits_{-\infty}
^{\infty}[\lambda^{1/2}- (\lambda-u_0(x'))^{1/2}]dx'=
\rho_-(\lambda)+\rho_+(\lambda) -i\tau(\lambda)\, .
\ee
Then one can observe
that the following identity holds
\be \label {f0}
\oint \limits _{a_{\infty}}f(\lambda)\Lambda_j=0 \, .
\ee
Really, $f(\lambda)\sim \lambda^{-1/2}$ and $\Lambda _j \sim \lambda ^{-3/2}
d\lambda$ as $\lambda \to \infty$.
Considering (\ref{f0}) as the integral over the contour surrounding the
interval $(0,1)$ we arrive at the identity
for analytic $\rho_-(\lambda)$, $\rho_+(\lambda)$ and
$\tau(\lambda)$.
\be \label{}
\oint \limits _{\cup\alpha_k} \{\rho_-(\lambda)+\rho_+(\lambda)\}\Lambda_j-
i\oint \limits _{\cup \beta_n}\tau(\lambda)\Lambda_j=0\, .
\ee

\section*{Figure Captions}

\begin{itemize}
\item{1. Ambiguity in local determination of genus.
Solid line: solution of the Whitham equations $(N=1)$,
broken line: solution of the Hopf equation $(N=0)$.}

\item{2. Separatrix $\{F(t)=\{ (x,r ): r(x,t)= r_1, \dots, r_{2N+1}\}\}$ ,
one-hump case. \\
a) $F(0)=\{(x,r): \ r=u_0(x)\}$  \ \
b) $F(t), \ \ t>t_{crit}$ }

\item{3. Multi-hump initial data. \ \
a) $\rho$- and $\tau$- domains. \ \
b) Sequence $\{x_{k_j}^\pm$\}. }

\end{itemize}

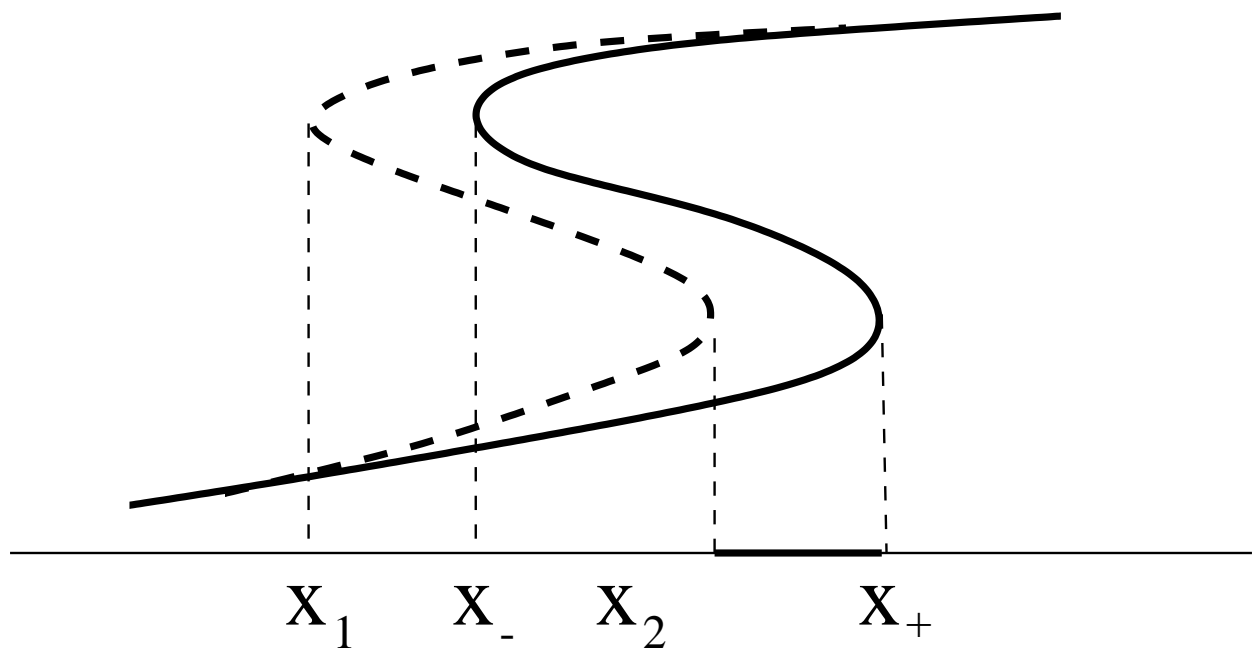
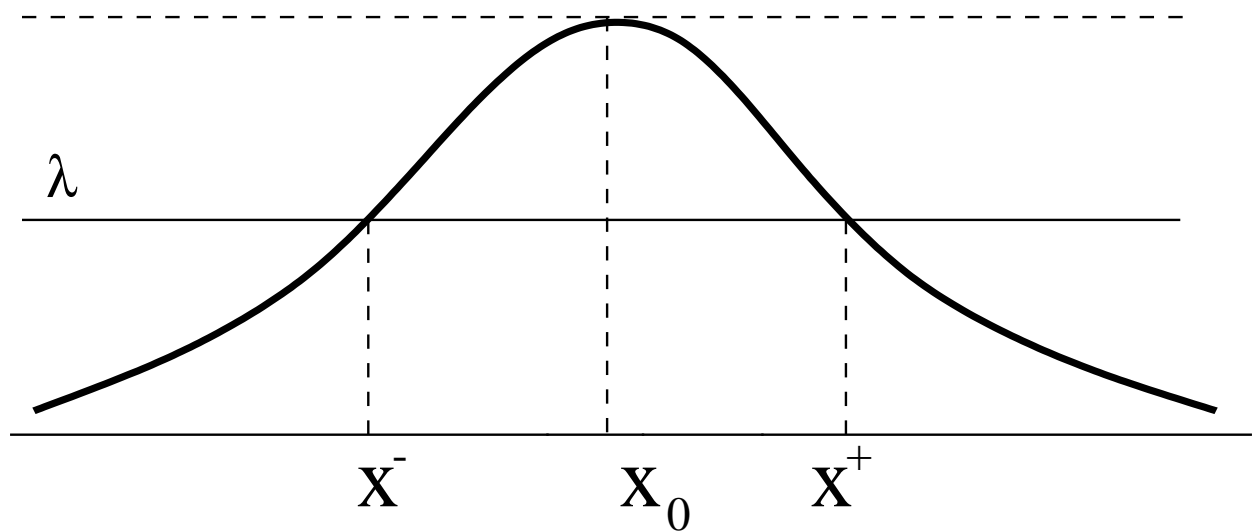
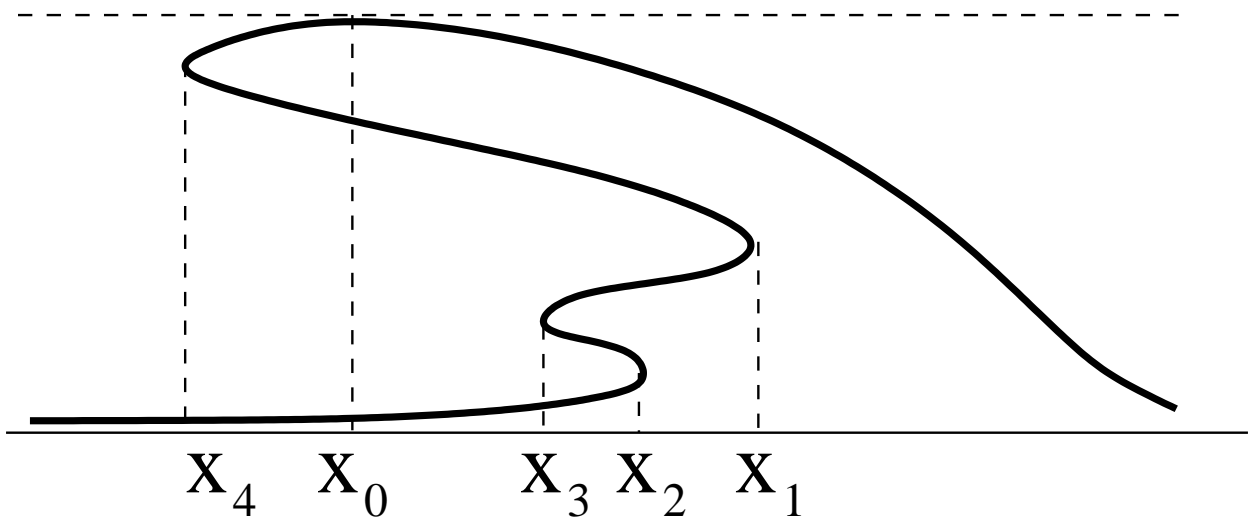
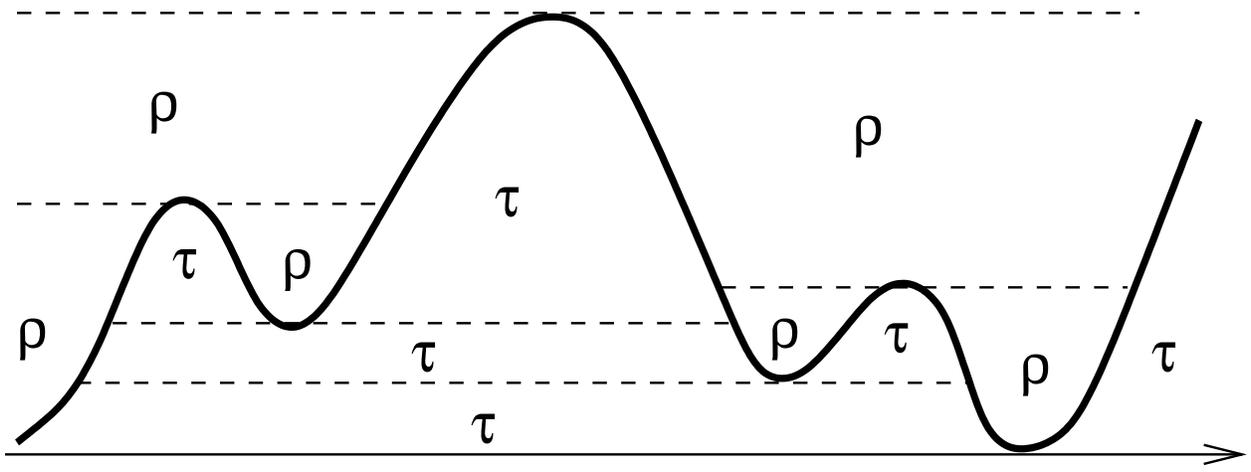
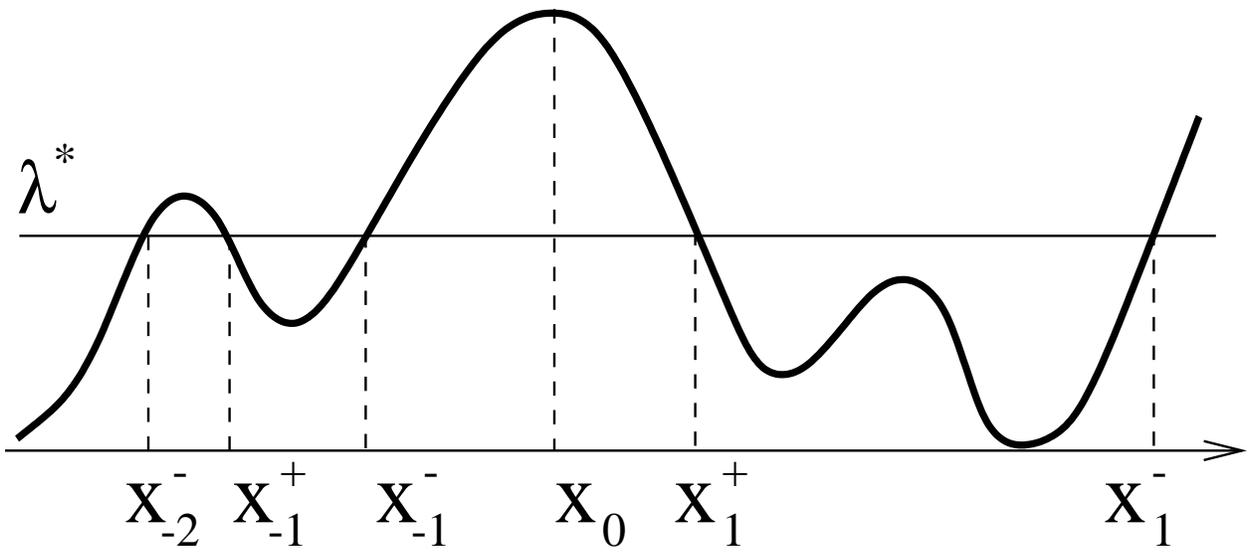
\end{document}